# Perovskite Twin Solar Device with Estimated 50% Bifacial PCE Potential and New Solar Material Options

Hans Hermann Otto
*TU Clausthal, D-38678 Clausthal-Zellerfeld, Germany*
hhermann.otto@web.de

**Abstract**

There are recent investigations regarding tandem solar cells with a top perovskite cell and a bottom silicon one with a potential of > 25% power conversion efficiency. Because of still high production costs of silicon cells it is believed that this tandem cell does not satisfy future requirements. Here the construction of a low-cost $FAPbI_3$ twin solar cell is proposed with assumed *PCE* of about 30%. With an ingenious rear illumination even 50% bifacial power conversion efficiency should be feasible. Importantly, a single twin cell can deliver the minimum potential difference of 1.7 V to conduct water splitting in practice. Avoiding additional electrodes, the twin cell device may be expanded by a large band gap sensitizer film at the mirror plan, derived from known electrically isolating phosphor storage materials that can capture otherwise wasted high-energetic radiation. Further, the substitution of $TiO_2$ (rutile) by ferroelectric and photo-catalytically active $Bi_2SiO_5$ with its comparable energy gap is suggested. In addition, $CuO_{1-x}$ could serve as new back electrode material due to its high electric conductivity. An environmentally benign $(Cs,FA)_2(Na,Cu,Ag)Bi(I,Br)_6$ elpasolite solar absorber material is discussed. Alternatively, pyroelectric hexagonal bismuth sulfide iodide with its complex superstructure may deliver a promising multiple band gap feature with suggested singlet fission capability as intrinsic property to overcome its lower efficiency.

## 1. Construction of a Twin Solar Cell

Recently, I proposed a twin solar cell device [1,2]. The realization of such device requires quite transparent materials both for light harvesting and for electrodes. Then the direct sun light may partly reach the bottom cell, which in turn also receives radiation from the back side. It should be mentioned that bifacial light capture to a certain extent is already realized in semitransparent dye cells (*Grätzel*, 2001) [3].
The back side illumination may be enhanced by mirrors or optical, perhaps luminescent fibers. If 80 % of the incident light is absorbed in the top cell, the proportionate contribution to the *PCE* of the hole device is assumed to be 68% from the top cell, 14% from the bottom cell sun-side illuminated and 15% from low-light back illumination in addition to further 3% from rear illumination of the top cell (see yellow arrows in Figure 1). With a top cell efficiency of about 20% (*Zhou et al.*, 2014 [4]; *Jeon et al.*, 2015 [5]; *Lin et al.*, 2015 [6]; *Yang et al.*, 2015 [7]) the overall twin cell efficiency would reach 30%. By an ingenious rear illumination technique this value may be even better and should reach 50%. The semi-transparency of the stacked device may be adapted by thickness variation of the active layers for optimum solar efficiency.

Importantly, a single twin cell can deliver 1.7 V minimum potential difference to conduct water splitting in practice (*Luo et al.*, 2014) [8].

A semi-transparent state of the art FAPbI$_3$ perovskite cell (Yang et al., 2015) [7] may be considered as backbone, but with transparent CuI as hole conducting layer.

CuI is a smart transparent p-type semiconductor with an experimental band gap of $E_g$= 3.1 eV, a dielectric constant of $\varepsilon = 6.5$ and a hole effective mass of $m_h^* = 0.38$ (*Zhu et al.*, 2012) [9] that guarantees high hole charge carrier transport. This was recently demonstrated by *Christians et al.* (2014) [10] using copper iodide on top of a hole transporting *PEDOT*:*PSS* layer to increase the open circuit voltage of a device.

The twin plane of the device is a mica sheet acting as *Kagomé* lattice substrate for the deposition of perfectly grown CuI films on both surfaces.

A (111)-oriented thin layer of CuI can be grown on a freshly cleaved muscovite sheet by way of thermal vacuum deposition of a (111)-oriented copper film and its subsequent transformation into CuI, when the copper is attacked by vapor from a weak solution of iodine. The cubic lattice parameter of copper is 3.615 Å, yielding a distance of $d_{(111)} = 2.56$ Å. This distance is in good agreement with the Cu-I bond length of $d_{Cu-I} = 2.62$ Å. A more expensive alloy of copper with some gold of composition Cu$_{0.805}$Au$_{0.195}$ is ideally adapted to the Cu-I distance, but would leave some gold on the mica surface, after copper was converted into iodide. The remaining gold will accumulate in nano-clusters due to its high surface mobility. As an optical meta-material these clusters do not lessen the transparency significantly. They even increase the conductivity and excellently cooperate with CuI as back electrode, extracting light induces holes. If an alloy instead of pure copper is vapor deposited, one should bear in mind that copper has a higher vapor pressure than gold.

Control of the layer structure of deposited CuI as a tetrahedrally bonded semiconductor with its tendency to layer disorder may be important to optimize its band gap and effective hole mass. These properties are influenced by the interstitial channel length of the polytype, as indicated in the case of SiC (*Matsushita and Oshiyama*, 2014) [11]. From this argument a wurtzite-type (*β*-CuI) deposit is favored. Also the net polarity of the deposited sheet caused by such wurtzite layers (polar space group *P6$_3$mc*) should be considered supporting charge separation when aligned in the optimum direction normal to the back electrode.

Both surfaces of the muscovite support will be coated with CuI resulting in the centerpiece of the twin cell. On both sides then a state-of the art MAPbI$_3$ or FAPbI$_3$ package will be assembled consisting of covering glass, *FTO* electrode, the hole depleting rutile/ZrO$_2$ buffer and the light harvesting perovskite film with (111)-preferred orientation (*Yang et al.*, 2015) [7]. Figure 1 shows the resulting solar cell device consisting of two cells mirrored at a transparent mica substrate ore fused silica one.

Because muscovite mica is a less disposable natural material, supporting the deposition of highly oriented CuI transparent films for quick research success, it may in future be replaced by transparent and less costly substrates. Mercapto-silane surface activated fused silica has been reported to strongly bind thin gold films deposited by e-beam or thermal evaporation (*Kossoy et al.*, 2014) [12]. The deposition of a copper bearing alloy discussed before may bind the better at the silica surface. Alternatively, avoiding vacuum technique, a thin CuI film can be deposited on a proper substrate like fused silica by a sol-gel process using acetonitrile-solved CuI and *TMED* as a stabilizer (*Zainun et al.*, 2011) [13].

Recently, a semi-transparent silver nanowire electrode has been used in a tandem solar cell device by *Bailie et al.* (2014) [14]. But silver is not considered as electrode material for the twin cell, because it can be attacked over time by iodine to form AgI. Also *PEDOT*:*PSS* provided as additional *HTL* is still harmful slowly corroding silver but not gold (*Suh et al.*, 2012) [15].

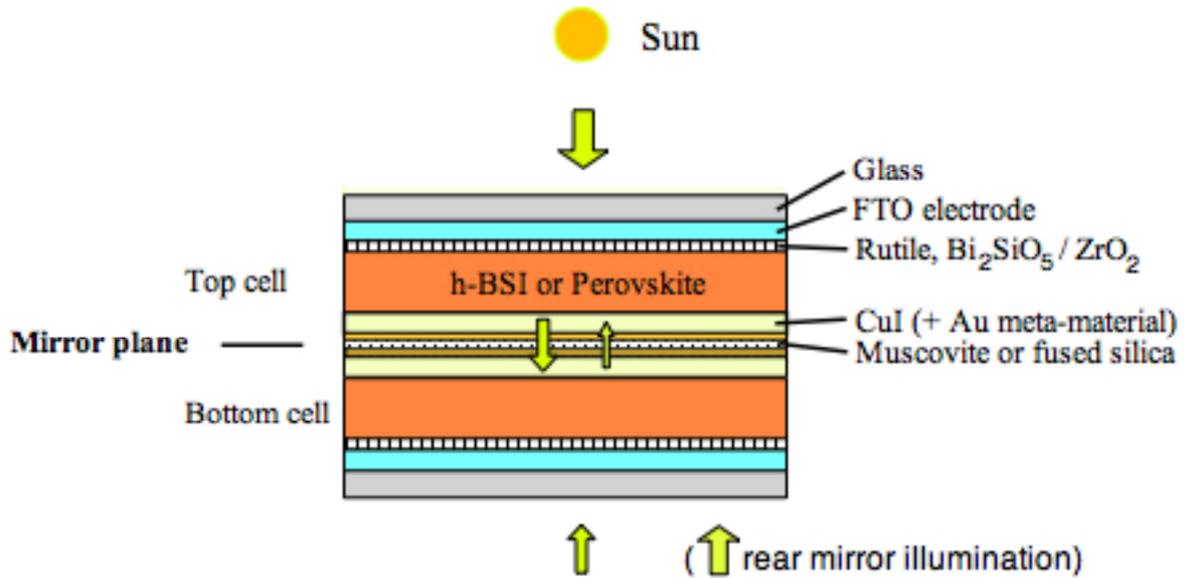

**Figure 1.** Perovskite twin solar cell as depicted in recent publications [1,2].

Basically, nearly perfect perovskite crystals may be formed using an elaborated growing technique without a seeding substrate on the twin plane. Then the silica (mica) support is dispensable. However, if one still uses two complete old-fashioned but quite transparent cells in a twinned arrangement, there is the additional possibility to deposit between their back electrodes a thin film of transparent and electrically isolating material of a very large band gap that are able to convert otherwise wasted high energy radiation into visible light. This light will shine in both cells and can enhance the solar efficiency of the device. The absorbed and dissipated energy from the photoelectric effect, *Compton* scattering or pair production finally creates photons in the desired low-energy spectral range. Candidate materials are derivatives of the known phosphor storage compounds such as $BaFBr:Eu^{2+}$ and $CsBr:Eu^{2+}$ with $Eu^{2+}$-$O^{2-}$ dipole hole traps [16-19]. The realization of such intensifier option, which does not need additional electrodes, depends on whether the energy gain is cost-effective.

## 2. Multiple Twin Solar Cell Stack

Multiple twinned structures are common features of ferroelastic compounds. A nice microscopic picture for such behavior of the ferroic anti-perovskite $Pb_3GeO_5$ was published years ago [20]. The concept of such structure may be applied to a solar device that could be produced in near future with 3d printing technique. If a triple stack is considered, the thickness of the photo-active layer may be chosen less than the carrier diffusion length of 1 µm to reduce the absorption pathways, which is important to light that strikes the surface under less than 90°. In addition, only single 'back' electrodes without fused silica seeding supports on the twin planes should be used. The more transparent the photo-active layer and the electrodes are, the higher the *PCE* of the new device will be. In addition, a rear illumination would be highly efficient and noticeably enhance the profit to cost ratio. The optimum alignment of the device along the position of the sun would be desirable, too. Using a *3d* printer, also the *p*-type and *n*-type electrode contacts could be guided out in alternating directions for purpose of wiring. A model calculation with given material transparencies and varied thicknesses of photo-active layers and electrodes will decide about the feasibility of triple stack or multiple stack solar cells and whether higher technical effort will be rewarded with a much greater power output. An example for such model calculation of a simple dye solar cell is given by the work of *Wenger et al*. [21]. The multiple stack approach is very interesting even for solar absorbers with a carrier diffusion length significantly smaller than 1 µm, because the absorber layer thickness can be adapted to the diffusion length to optimize the *PCE*, whilst the entire absorber volume needs not be reduced.

## 3. Importance of a Ferroic and Nano-Sized Conductive Domain Structure

The very high *PCE* of organic lead halide perovskites is the result of enhanced charge separation and improved carrier lifetimes with diffusion lengths > 1µm, favored according to *Frost et al.* [22, 23] by a small-sized ferroelectric domain structure, which in turn is caused by a low and less confining *Madelung* potential due to the low mean cationic charge of $<q>_c$ = $1.5^+$. However, others reported the non-ferroelectric nature of the conductance hysteresis, for instance of MAPbI$_3$, confirming the *I4/mcm* space group [24]. A careful assessment of experimental displacement ellipsoids may confirm conclusively whether this space group is correct. Recently, conductivity by both ions and electrons has been evidenced in an excellent study [25]. Further, a ferroelastic domain pattern was observed for MAPbI$_{3-x}$Cl$_x$ as interesting result of a piezoresponse force microscopic study [26]. Ferroelastic interactions are known to be very strong, and ferroelastic domains evidently improve photochemical reactivity and local separation of photo-generated carriers [27]. Ferroelastic forces causing nanodomain formation will strongly influence electric conductivity and pathways for charge transport. Anyway, a ferroic and highly conductive nano-sized domain structure is the key to the high performance of perovskite solar cells. Optimizing the ferroelastic domain structure by chemical substitutions will result in a maximum solar efficiency. Notably, the recent observation of stable room-temperature ferroelectricity in strain-free ultra-thin SrTiO$_3$ films, reported otherwise non-ferroelectric, can open new routes of materials adjustment in nano-scale solar devices [28].

## 4. Comparison with High-$T_c$ Superconducting Oxide Perovskites

A less restricted freedom of electronic movement may also be suggested for the family of superconductors related to oxide perovskites. The rise of the transition temperature $T_c$ goes inversely proportional with a power of the mean cationic charge $<q>_c$ [29,30]. The step by step lowering of the *Madelung* energy allows the formation a ferroelastic and conductive domain structure of ever narrower size resulting in a gradual increase of the transition temperature. Surprisingly, again $<q>_c \approx 1.5^+$ emerges as the high-$T_c$ asymptotic limit. The empiric relation of $T_c$ versus $2740 \cdot <q_c>^{-4}$ points the way to compounds with a superconducting transition above room temperature. Empirically, the connection of *Fibonacci* numbers with assumed domain widths may evidence the chaotic-filamentary nature of high-$T_c$ superconductivity [30].

## 5. Recommended New Materials

### 5.1 Bi$_2$(Si,Ge)O$_5$ (blocking layer)

The most widely used blocking layer to prevent charge recombination losses in perovskite solar cells is composed of nanorod straw crystals of rutile (TiO$_2$), which show a surface defect dipole structure responsible for photo-active response [31]. Doping can enhance this property as demonstrated recently [32].
However, having in mind the amplifying effect of ferroic properties on the solar *PCE*, it is recommended to replace rutile (TiO$_2$) by ferroelectric and photo-catalytically active Bi$_2$SiO$_5$ with its comparable energy gap of $E_g$= 3.2 eV and large spontaneous polarization of $P_c$ = 23.5 µC/cm$^2$ [33, 34]. The calculated mean refractivity index yields $n$ = 2.38. The crystal structure is acentric with space group *Cc*, and lattice parameters of $a$ = 15.1193(1) Å, $b$ = 5.4435(1) Å, $c$ = 5.2892(1) Å, and $\beta$ = 90.070(2)°, respectively.

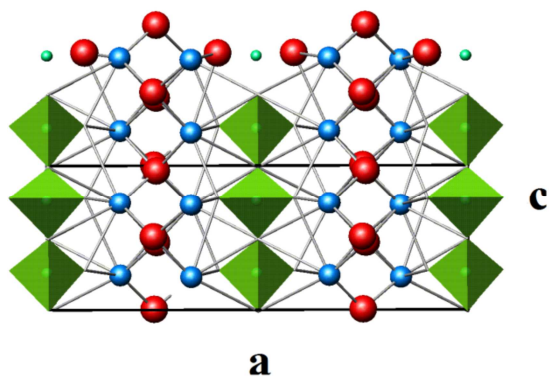

**Figure 2**. Crystal structure of $Bi_2SiO_5$ projected down (010). Twisting silicate tetrahedral chains (light green) running along the *c* axis (direction of main ferroelectric polarization). Bi blue, oxygen not bound to silicate in red color.

Basically, this structure is composed of twisted $SiO_3^{2-}$ silicate tetrahedral chains running down the *c* axis. The chains are separated by $Bi_2O_2^{2+}$ *Aurivillius* sheets [35]. The crystal structure is depicted in figure 2. This environmentally benign material option may also support charge separation in solar cells. Apart from the blocking layer option, doping or chemical substitution may deliver a very interesting solar energy harvesting material. Best suitable would be a partial replacement of the $Bi_2O_2^{2+}$ sheet by $CsBiI_2^{2+}$ or $CsBiBr_2^{2+}$, thereby converting the insulator to a semiconductor and shifting the energy gap to lower energy.

**5.2 $CuO_{1-x}$ (back electrode material)**

Besides CuI, an interesting electrode material that needs further investigation for solar cell application is oxygen depleted cupric oxide $CuO_{1-x}$ as a hole conductor. Incidentally, this phase is obviously responsible for filamentary superconductivity at a very high transition temperature of $T_c$ = 220 K, when formed as part of a multiphase $YBa_2Cu_3O_7$ thin film sample [36]. Subsequently, *Osipov et al.* [37] observed *HTSC*-like giant electric conductivity at an interfacial layer formed between CuO and Cu even above 300 K. This temperature region would be required for solar cell application. The electric conductivity exceeds enormously the reference value for Cu by a factor of $1.5 \cdot 10^5$. Filamentary superconductivity may be assumed to explain this observation, too. However, the higher light absorbance in contrast to CuI, for instance, may hinder its application as electrode material in a twin cell device unless the layer can be designed extremely thin.

**5.3 Less Toxic Bismuth Compounds Instead of Lead Perovskite Absorbers**

A major drawback of the recommended $PbI_2$-based perovskites is the toxicity of lead. Therefore, the emphasis will be placed here on some less toxic *n*-type bismuth compounds now under investigation to construct a completely inorganic device.
Before proposing bismuth compounds as solar absorbers, one should check the availability of the metal, because toxic lead will be replaced more and more by environmentally benign bismuth, for instance in solders and specialty alloys. Besides pure bismuth sulfosalt minerals such as bismuthinite ($Bi_2S_3$) and galenobismuthinite ($PbBi_2S_4$), a sufficient amount of Bi is found as $Bi_2S_3$ solid solution in the most abundant lead sulfide ore galenite (PbS), where $Bi_2\square S_3$ replaces 3PbS via formation of vacant metal positions [38].

### 5.3.1 Elpasolites

Elpasolites are ordered double perovskites such as synthetic $Cs_2NaBiI_6$ with differently charged octahedrally coordinated cations [39]. A further example is $Cs_2(Li,Na)La(Br,I)_6$. Large single crystals have been grown and used as scintillators [40]. According to chapter 3, the mean cationic charge of $<q_c> = 1.5$ for such halogenides suggests low cohesive energy with high *PCE*.
In the following, a Bi-based $(Cs,FA)_2(Na,Cu,Ag)Bi(I,Br)_6$ elpasolite is proposed as potential solar absorber. Despite the fact that CuI and AgI like tetrahedral coordination under ambient conditions, the substitution of octahedrally coordinated $Na^{1+}$ by $Cu^{1+}$ or $Ag^{1+}$ may be recommended to allow (virtual) charge transfer between their 1+ and 2+ oxidation states. Replacement of $I^{1-}$ by $Br^{1-}$ should force at least $Ag^{1+}$ more readily in an octahedral environment of bromine ions, because pure AgBr shows the rocksalt structure. Also $Bi^{3+}$ allows virtual charge transfer between its 3+ and 5+ oxidation state. This is important for the charge carrier generation process. Further, the substitution of some $Cs^{1+}$ by the larger organic $FA^{1+}$ cation should stabilize a non-cubic crystal structure. Such mixed-cation substitution was recently applied by *McMeekin et al.* [41]. Deviation from the cubic symmetry is desirable, contrary to cubic scintillator materials for gamma ray and neutron detection, so that a ferroelastic domain structure can be formed to open effective pathways for rapid charge carrier transport. In addition, doping with $Eu^{2+}$–$O^{2-}$ can form traps supporting the conversion of otherwise wasted ultraviolet radiation into charge carriers.

### 5.3.2 Hexagonal Bismuth Sulfide Iodide

Recently, the superstructure and some properties of hexagonal bismuth sulfide iodide (hBSI) were published [1]. Nano-rods of this compound can be epitaxially grown onto a (111) CuI layer supported by some lattice matches. The favorable needle-like morphology is caused by the small energetic period along the *c*-axis, comparable to the straw-crystal habit of rutile, $TiO_2$. Again, with its *c*-axis in the favorable polarity direction aligned normal to the electrode plane, the morphology and pyroelectric response of the compound may help to separate light induced electron carriers from hole ones with a higher throughput.
$Bi_{5-x}(Bi_2S_3)_{39}I_{12}S$ exhibits a complex *Fibonacci* superstructure with some vacant metal positions and empty channels along 0,0,z [1]. It should be investigated, whether this structural details show singlet fission capability due to multiple mutually separate band gaps. Whereas efficient intramolecular singlet fission, the carrier multiplication due to formation of two triplet excitons via splitting of a singlet exciton, was observed in organic materials such as pentacene [42], inorganic materials are still being sought. The energy gap of hexagonal bismuth sulfide iodide of $E_g \approx 1.6$ eV is comparable with that of $MAPbI_3$ ($E_g = 1.55$ eV).
Besides the synthesis route by epitaxy described previously [1], a low temperature synthesis can be carried out by means of diluted aqueous solutions containing stoichiometric amounts of $Bi(NO_3)_3$ (acidified with nitric acid), thiourea and hexadecyl-trimethyl-ammonium iodide, respectively, exposed to super-cavitation by ultrasonic waves to precipitate a dense hBSI thin film on a substrate or back electrode with subsequent washing out of not wanted byproducts and heat treatment at 80°C, using a modified method of *Deng et al.* [43].

## 6. Outlook

In future, the environmentally friendly and at the same time strategic tool of decentralized energy management, which avoids long energy transport routes, should combine the high electric energy production possible by the twin solar technique with local energy storage in recently developed rechargeable Li-$O_2$ batteries, which cycle via LiOH formation and decomposition, already charging about 5.7 kWh per kg [44].

## 7. Conclusions

The proposed twin solar device is able to absorb sunlight from both the front and the rear side, giving much more bifacial power conversion efficiency (*BPCE*) up to 50%. The development of a FAPbI$_3$ perovskite twin solar cell is desirable. Notably, the low mean cationic charge of <$q_c$> = 1.5$^+$ of organic inorganic halide perovskites is equivalent to a reduced *Madelung* energy, which obviously favors high conductivity and solar efficiency due to the formation of a narrow domain structure of possibly ferroelastic nature as charge transport pathway for both ions and electrons. A new bismuth silicate (germanate) blocking layer material is recommended to replace rutile. In addition, CuO$_{1-x}$ may serve as new extremely conductive back electrode material. A less toxic (Cs,FA)$_2$(Na,Cu,Ag)Bi(I,Br)$_6$ elpasolite material is discussed as promising new solar absorber. Alternatively, but expected to be less efficient, the less toxic hexagonal bismuth sulfide iodide may be a material option. This sustainable material should be investigated in more detail to unravel its possible singlet fission capability due to intrinsic vacant positions and empty channels in its crystal structure, leading as assumed to multiple mutually separate band gaps. Right now, available Si-based cells, CdTe and CIGS thin film ones, respectively, should still be used, but in a twinned arrangement, whenever bifacial lighting conditions are given, because we are reaching at some point the limits of what can built over with solar panels.

## References

[1] H. H. Otto (2015). *World J. of Cond. Matter Phys.* **5,** 66-77.
[2] H. H. Otto (2015). *PSCO-2015 conference, Lausanne, Book of Abstracts,* p.183.
[3] M. Grätzel (2001). *Nature* **414**, 338-344.
[4] H. Zhou et al. (2014). *Science* **345**, 542-546.
[5] N. J. Jeon et al., (2015). *Nature* **517**, 476-480.
[6] Q. Lin et al., (2015). *Nature Photonics* **9**, 106-112.
[7] W. S. Yang et al. (2015). *Science* **348**, 1234-1237.
[8] J. Luo et al. (2014). *Science* **345**, 1543-1546.
[9] J. Zhu, R. Pandley and M. Gu (2012). *J. Phys. Cond. Mater.* **24,** 475503-475511.
[10] J. A. Christians et al. (2014). *J. Am. Chem. Soc.* **136**, 758-764.
[11] Y. Matsushita and A. Oshiyama (2014). *Phys. Rev. Lett.* **112**, 136403, 1-5.
[12] A. Kossoy et al. (2015). *Adv. Optical Mater.* **3**, 71-77.
[13] A. R. Zainun, U. M. Noor and M. Ruso (2011). *Int. J. Phys. Sci.* **6**, 3993-3998.
[14] C. D. Bailie et al. (2014). *Energy Environm. Sci.* DOI: 10.1039/c4ee03322a
[15] Y. Suh et al. (2012). *Appl. Mater. Interfaces* **4**, 5118-5124.
[16] V. V. Mikhaylin and M. A. Terekhin (1989). *Nucl. Instrum. Methods* **60**, 2545.
[17] E. Radzhabov and T. Kurobori (2001). *J. Phys. Condens. Matter* **13**, 1159-1169.
[18] S. Hesse et al. (2009). *Radiat. Meas.* **42**, 638-643.
[19] P. Leblans, D. Vandenbroucke and P. Willems (2011). *Materials* **4**, 1034-1086.
[20] H. H. Otto (1979). *Z. Kristallogr.* **149**, 227-240.
[21] S. Wenger et al. (2009). *24. Europ. Photovolt. Solar Energy Conf. and Exhib.*, Hamburg.
[22] J. M. Frost, K. T. Butler and A. Walsh (2014). *APL Mat.* 2, 081506.
[23] J. M. Frost et al. (2014). *Nano Lett.* **14**, 2584-2590.

[24] J. Beilstein-Edmands et al. (2015). *arXiv*: 1504.05454v1 [cond-mat.mtrl-sci].
[25] C. Eames et al. (2015). *Nature Comm.* **6,** Article number 7497, 1-15.
[26] I. M. Hermes et al. (2015). *PSCO-2015 conf.*, *Lausanne, Book of Abstracts*, Poster 43.
[27] R. Munprom, P. A. Salvador and G. S. Rohrer. *J. Mater. Chem.* A, 2016, Advance Article.
[28] D. Lee et al. (2015). *Science* **349**, 1314-1317.
[29] H. H. Otto (2008). *arXiv*: 0810.3501v1 [cond-mat.supr.-con].
[30] H. H. Otto (2015). Superconductivity–Playing with Numbers. To be published in *WJCMP*.
[31] G. Li et al. (2004). *Appl. Physics Lett.* **85**, 2059-2061.
[32] D. Wojcieszak et al. (2014). *Int. J. Photoenergy* 2014, Art. Id 463034, 10 pages.
[33] H. Taniguchi et al. (2013). *Angew. Chem. Int. Ed.* **52**, 8088-8092.
[34] Y. Kim et al. (2014). *IUCrJ research letters* **1**, 160-164.
[35] B. Aurivillius (1949). *Ark. Kemi*, p. 463.
[36] R. Schönberger et al. (1991). *Physica* **C173**, 159-162.
[37] V. V. Osipov et al. (20010). *J. Exp. Theor. Phys.* **93**, 1082-1090.
[38] H. H. Otto and H. Strunz (1968). *N. Jb. Miner. Abh.* **108**, 1-19.
[39] L. R. Morss and W. R. Robinson (1972). *Acta Crystallogr.* **B28**, 653.
[40] P. Yang et al. (2009). *MRS Proc.* **1164**, L11-05
[41] D. P. McMeekin et al. (2016). *Science* **351**, 151-155.
[42] E. Busby et al. (2015). *Nature Materials* **14**, 426-433.
[43] C. Deng et al. 2013). *Mater. Lett*. **108**, 17-20.
[44] T. Liu et al. (2015). *Science* 350, 530-533.